# Motion of Planetesimals in the Hill Sphere
# of the Star Proxima Centauri

## S. I. Ipatov*

*Vernadsky Institute of Geochemistry and Analytical Chemistry, Russian Academy of Sciences, Moscow, Russia*
*\*e-mail: siipatov@hotmail.com*

**Abstract**—The motion of planetesimals initially located in the feeding zone of the planet Proxima Centauri $c$, at distances of 500 AU from the star to the star's Hill sphere radius of 1200 AU was considered. In the analyzed non-gaseous model, the primary ejection of planetesimals from most of the feeding zone of an almost formed planet $c$ to distances greater than 500 AU from the star occurred during the first 10 million years. Only for planetesimals originally located at the edges of the planet's feeding zone, the fraction of planetesimals that first reached 500 AU over the time greater than 10 million years was more than half. Some planetesimals could reach the outer part of the star's Hill sphere over hundreds of millions of years. Approximately 90% of the planetesimals that first reached 500 AU from Proxima Centauri first reached 1200 AU from the star in less than 1 million years, given the current mass of the planet $c$. No more than 2% of planetesimals with aphelion orbital distances between 500 and 1200 AU followed such orbits for more than 10 million years (but less than a few tens of millions of years). With a planet mass equal to half the mass of the planet $c$, approximately 70–80% of planetesimals increased their maximum distances from the star from 500 to 1200 AU in less than 1 million years. For planetesimals that first reached 500 AU from the star under the current mass of the planet $c$, the fraction of planetesimals with orbital eccentricities greater than 1 was 0.05 and 0.1 for the initial eccentricities of their orbits $e_o = 0.02$ and $e_o = 0.15$, respectively. Among the planetesimals that first reached 1200 AU from the star, this fraction was approximately 0.3 for both $e_o$ values. The minimum eccentricity values for planetesimals that reached 500 and 1200 AU from the star were 0.992 and 0.995, respectively. In the considered model, the disk of planetesimals in the outer part of the star's Hill sphere was rather flat. Inclinations $i$ of the orbits for more than 80% of the planetesimals that first reached 500 or 1200 AU from the star did not exceed 10°. With the current mass of the planet $c$, the percentage of such planetesimals with $i > 20°$ did not exceed 1% in all calculation variants. The results may be of interest for understanding the motion of bodies in other exoplanetary systems, especially those with a single dominant planet. They can be used to provide the initial data for models of the evolution of the disk of bodies in the outer part of Proxima Centauri's Hill sphere, which take into account gravitational interactions and collisions between bodies, as well as the influence of other stars. The strongly inclined orbits of bodies in the outer part of Proxima Centauri's Hill sphere can primarily result from bodies that entered the Hill sphere from outside. The radius of Proxima Centauri's Hill sphere is an order of magnitude smaller than the radius of the outer boundary of the Hills cloud in the Solar System and two orders of magnitude smaller than the radius of the Sun's Hill sphere. Therefore, it is difficult to expect the existence of a similarly massive cloud around this star as the Oort cloud around the Sun.

## INTRODUCTION

Recently, significant attention has been devoted to the study of exoplanetary systems. Reviews on this topic have been provided, in particular, by Marov and Shevchenko (2020, 2022). The Proxima Centauri system consists of a star with a mass equal to 0.122 solar masses and three planets. Star Proxima Centauri (α Centauri C) is a member of a triple star system, which also includes the binary star Alpha Centauri AB, consisting of two stars with masses close to the mass of the Sun. The distance between Proxima Centauri and Alpha Centauri AB is approximately 12950 AU. The mass of the largest planet Proxima Centauri $c$ is $m_c = 7m_E$ (where $m_E$ is the mass of the Earth). The semimajor axis of its orbit is $a_c = 1.489$ AU. Two other planets, Proxima Centauri $b$ and $d$, have smaller masses and semimajor axes of the orbits: $m_b = 1.17m_E$, $a_b = 0.04875$ AU, $a_d = 0.02895$ AU and $m_d = 0.29 m_E$.

Schwarz et al. (2018) studied the motion of exocomets in the Proxima Centauri system with initial orbital inclinations $i$ distributed as an Oort cloud with $0° \le i \le 180°$. The perihelion distances of the exocomet orbits were less than 0.0485 AU, and the initial eccentricities of the orbits ranged from 0.95 to 0.9999. The initial values of the semimajor axes of the exocomet orbits varied in different series of calculations from 1 to 1000 AU. Schwarz et al. (2018) considered the semimajor axis of the orbit of the planet Proxima Centauri b equal to 0.0485 AU, and the values of the semimajor axis of the orbit of the planet Proxima Centauri c were taken in the range from 0.06 to 0.3 AU (up to 0.7 AU for test calculations). It was later discovered that the semimajor axis of the planet c is much larger than previously assumed in the calculations by Schwarz et al. (2018). In these calculations, the probabilities of exocomet collisions with planets were approximately the same for the model that considered only the gravitational influence of the star and planets and the model that additionally took into account the gravitational influence of the binary star Alpha Centauri AB. For the latter model, Schwarz et al. (2018) found that within 2 Myr, more than 3/5 of exocomets with semimajor axes of initial orbits between 990 and 1000 AU either reached 20 000 AU from Proxima Centauri or collided with it.

This study uses the same results of modeling the evolution of the orbits of planetesimals that started from the vicinity of the orbit of Proxima Centauri c as (Ipatov, 2021, 2022, 2023a, 2023b). It is the final work in this series of calculations. In (Ipatov, 2023a), I considered the size of the feeding zone of the planet c (including the study of stable orbits within this feeding zone), while in (Ipatov, 2021, 2022, 2023b), the delivery of planetesimals from this feeding zone to the inner planets was estimated. Those studies primarily focused on the motion of planetesimals at distances less than 2 AU from the star. Below I study the motion of planetesimals from the same feeding zone at distances greater than 500 AU. The initial data in all calculations were the same; only the motion of the same initial planetesimals at different distances from the star was considered. The initial eccentricities of the planetesimal orbits equaled to 0.02 or 0.15. These calculations took into account the gravitational influence of the star and planets c and b without the effects of gas. For the given initial data, it was considered unimportant how relatively large planetary embryos formed. In a number of studies (e.g., Lambrechts and Johansen, 2012, 2014; Morbidelli, 2020; Morbidelli et al., 2015; Wahlberg Jansson K. and Johansen, 2014), pebble accretion is believed to play a significant role in the presence of gas in the disk. In our calculations, it was assumed that the orbit of planet c is beyond the snow line, and the orbit of planet b may be within the habitable zone. Ipatov (2021, 2022, 2023b) investigated the delivery of icy planetesimals from the feeding zone of the planet Proxima Centauri c to the planets Proxima Centauri b and d. Probability $p_b$ of a collision of a planetesimal from the feeding zone of planet c with planet b was estimated to be ~$2 \times 10^{-4}$ and $10^{-3}$ for initial eccentricities of planetesimal orbits of 0.02 and 0.15, respectively. The probability of a collision of a planetesimal with the star was approximately the same, while the probability of collision with planet d was likely lower than $p_b$ by no more than a factor of two. The probability of collisions of bodies migrating from the zones of giant planets with the Earth is considered to be lower than the above values and in most variants of calculations did not exceed $10^{-5}$ (Ipatov, 2020). The total mass of planetesimals delivered to planet b from the feeding zone of planet c, was estimated in the range of $2 \times 10^{-3} m_E$ to $1.5 \times 10^{-2} m_E$. With water content in planetesimals ranging from 0.05 to 0.5, the mass of water delivered to planet b, was estimated to be approximately $10^{-4} m_E$ to $10^{-2} m_E$ and probably exceeded the mass of water in the Earth's oceans ($2 \times 10^{-4} m_E$). The number of planetesimals reaching distances from the star greater than 500 AU was no less than the number of planetesimals that collided with the planets.

The Hill sphere of the star Proxima Centauri (relative to the binary star Alpha Centauri) is 1200 AU (Schwarz et al., 2018). Since Proxima Centauri is member of a triple star system, the radius of its Hill sphere is two orders of magnitude smaller than that of the Solar System. According to Chebotarev (1965), the radius of the Hill sphere of the Solar System is 230000 AU. Souami et al. (2020) estimate this radius at 0.865 pc, i.e., 178 419 AU. In this study, it was assumed that the Sun moves in a circular orbit with a semimajor axis of 25 000 light years, and the mass of the Galactic stellar disk is $2.32 \times 10^{11}$ solar masses. These authors also considered other spheres (including Chebotarev's and Laplace's spheres). Inside the Hill sphere lies the Oort cloud. The distance from the Sun to the lower and upper boundaries of the Oort cloud is considered to be about $10^3$ and $10^5$ AU, respectively.

In this paper, I first review publications on the formation of the Oort cloud in the Solar System. Further, I discuss the results of calculations regarding the motion of planetesimals in the outer part of Proxima Centauri's Hill sphere. The non-gaseous stage of formation of the planetary system is considered. The timescales over which planetesimals first reached 500 AU from the star are examined. The distribution of planetesimals that first reached 500 and 1200 AU from the star over their orbital eccentricities and inclinations is given. We also consider the time required for planetesimals' maximum distances from the star to increase from 500 to 1200 AU. The number of planetesimals that could be located at distances from the star between 500 and 1200 AU at various moments of time is estimated. The possibility of the formation of analogs of the Hills and Oort clouds in the Proxima Centauri system is discussed.

## HILLS AND OORT CLOUDS IN THE SOLAR SYSTEM

The Oort Cloud is considered to be the source of many comets (Emel'yanenko et al., 2018; Dones et al.,

2004, 2015; Fouchard et al., 2020; Emel'yanenko et al., 2005, 2007, 2013). According to Emel'yanenko et al. (2013), the observed comets of the Halley type, almost half of the comets of the Halley family, and more than 90% of the centaurs (for which $5 < q < 28$ AU and $a < 1000$ AU, where $q$ and $a$ are the perihelion distance and the semimajor axis of the orbit, respectively) came from the Oort cloud. The inner (flattened) part of the Oort cloud (from 1000–5000 to 20000 AU) is called the Hills cloud (Hills, 1981). Some scientists call the Oort cloud only its outer spherical part (from 20000 to 50000–200000 AU). Estimates of the boundaries of the Oort cloud can differ significantly among different authors. According to Fouchard et al. (2020), the Oort cloud is rather flat for $a < 6000$ AU, and isotropic for $a > 8000$ AU. Morbidelli (2005) estimated the mass of the Oort cloud to be equal to 3 $m_E$. Fernandez and Brunini (2000) suggested that the mass of the Oort cloud is several times that of the Earth, with most of this mass concentrated in the inner core of the cloud.

Several models have been proposed for the formation of the Oort cloud. The "classical" model considers the ejection of bodies from the zone of the giant planets during the formation of the Solar System (Safronov, 1972). Duncan et al. (1987) started their calculations with $a = 2000$ AU, orbital inclinations $i$ equal to 18°, and perihelion distances from 5 to 35 AU. Their calculations took into account perturbations from planets, stars and galaxies. The integration was carried out over a time interval of 4.5 Gyr. It was concluded that the inner part of the Oort cloud ($a < 20000$ AU) contains approximately five times more comets than at $a > 20\,000$ AU.

In (Dones et al., 2004), the semimajor axes of the orbits of the initial bodies ranged from 4 to 40 AU, and their eccentricities and inclinations were small. It was found that after 4 Gyr, approximately 2.5% of the matter ejected from the Solar System remained in the Oort cloud. Morbidelli (2005), objecting to this model, noted that such a model would require the mass of the initial planetesimals in the zone of Uranus and Neptune to be about $100m_E$, while the Nice model (Brasser and Morbidelli, 2013; Clement et al., 2018, 2019; Gomes et al., 2005; Levison et al., 2011; Morbidelli et al., 2005, 2010; Tsiganis et al., 2005) needs a smaller mass of planetesimals for the migration of the embryos of Uranus and Neptune to their current orbits. In the Nice model, the trigger for sharp changes in the orbits of these embryos is the capture of Jupiter and Saturn into the 1 : 2 orbital resonance. In my opinion, this objection by Morbidelli (2005) may be insignificant, since the calculations of Ipatov (1991b, 1993, 2000) demonstrated that the migration of the embryos of Uranus and Neptune from the orbit of Saturn to their current orbits is achievable with a total mass of planetesimals in the zone of the giant planets from 135 $m_E$ to 180 $m_E$. In these calculations (Ipatov, 1991a, 1991b, 1993, 2000), the capture of Jupiter and Saturn into resonance was not considered, and the embryos of Uranus and Neptune migrated to their current orbits solely due to gravitational interactions with planetesimals. More than 80% of the planetesimals were ejected to distances greater than 500 AU from the star. It was concluded in (Ipatov, 2000) that a disk of planetesimals with a mass of $100m_E$ is sufficient for the migration of Uranus and Neptune embryos to their current orbits. This mass is lower if we consider larger (than in the calculations) semimajor axes of the initial orbits of Uranus and Neptune embryos (in the calculations they were 8 and 10 AU). The main changes in the orbital elements of the giant planet embryos in the calculations of Ipatov (1993, 2000) occurred over a period of no more than 10 million years, although individual bodies could collide with these embryos over the course of billions of years. The Nice and Ipatov models are also discussed in (Marov and Ipatov, 2023). A large total mass of planetesimals in the zone of the giant planets was also considered by Safronov (1972). He estimated the mass of a protoplanetary cloud to be 0.05 solar masses. With dust fraction of 0.015 in the cloud, this cloud mass corresponds to a solid substance mass of $250m_E$. The latter estimate can be up to five times larger than the mass of solid matter in the planets. Such an estimate of the ratio of the mass of ejected planetesimals to the mass of solid matter incorporated into the planets is consistent with the results of planetary accretion calculations (Ipatov, 1987, 1993, 2000). Even before the discovery of the first trans-Neptunian objects, in (Ipatov, 1987), in particular, it was noted that some trans-Neptunian objects could have come from the feeding zone of the giant planets, while others could have formed beyond the orbit of Neptune. Ipatov (2000) pointed out that Jupiter ejected bodies into more eccentric orbits than Uranus and Neptune. Safronov (1972) assumed that 5% of the bodies ejected from planetary feeding zones remained in the Oort cloud.

Dybczynski et al. (2008) studied the orbital evolution of bodies that were initially in almost circular coplanar orbits at a distance of 4 to 50 AU from the Sun. In their calculations, only 0.3% of the particles remained in the outer part of the Oort cloud (with $q > 45$ AU and $a > 25000$ AU) after 1 Gyr. Emel'yanenko et al. (2007, 2013) considered a model similar to (Dones et al., 2004), but for computational efficiency, bodies that were initially in eccentric orbits were considered. The perihelia $q$ of the initial orbits ranged from 5 to 36 AU, with $50 < a < 300$ AU. In (Emel'yanenko et al., 2007) after 4.5 Gyr, 17% of the bodies had $a > 10^3$ AU, and 9.4% of the bodies had $a > 10^4$ AU.

It is believed that during the formation of the Solar System there were more stars in its vicinity than now (Brasser et al., 2006; Fernandez and Brunini, 2000). Fernandez and Brunini (2000) considered the formation of the Oort cloud in a denser stellar environment than at present. In their calculations, the initial values of the semimajor axes of the bodies' orbits were 100 or 250 AU, and the perihelia of their orbits were in the range from 4 to 30 AU. The initial orbital inclinations were 0.1 or 0.2 rad. The time interval of 100 million years was considered. Gravitational interactions with planets were taken into account using the spheres' method. It has been found that in this case, the number of bodies in the outer regions of the Oort cloud is much smaller than with the formation of the Oort cloud in the present stellar environment. According to Brasser et al. (2006), the model of the formation of the Oort cloud in a denser stellar environment can explain the orbits of bodies like (90377) Sedna. In their calculations, they examined the initial orbits of bodies with $4 < a < 12$ AU and $20 < a < 50$ AU. The model included gravitational effects from the Sun, Jupiter, and Saturn. The time interval of 3 million years was considered. A symplectic integrator was used to solve the equations of motion of bodies (Levison and Duncan, 1994). Brasser et al. (2010) calculated the formation of the Oort cloud at various distances of the Solar System from the center of the galaxy.

Levison et al. (2010) believed that about 90% of the Oort cloud bodies are of extrasolar origin. Siraj and Loeb (2021) also suggested that interstellar objects dominate in the Oort cloud, and their fraction increases with greater distance from the Sun.

## CALCULATION VARIANTS

In the considered calculation variants, the evolution of the orbits of planetesimals from the feeding zone of the Proxima Centauri planet $c$ was studied. When modeling the motion of planetesimals, the gravitational influence of the star and planets $c$ and $b$ was taken into account. In a number of variants, in addition to calculations with the present mass of planet $c$, $m_c = 7m_E$, calculations were carried out for the mass of the planet moving in the orbit of planet $c$ equal to $k_c = 0.5$ or $k_c = 0.1$ of its present mass. The orbital inclinations of the planets were taken equal to zero. The initial eccentricities of the planetesimal orbits were $e_o = 0.02$ or $e_0 = 0.15$. The initial inclinations of the planetesimal orbits were equal to $e_o/2$ rad (i.e., $0.57°$ or $4.3°$ for $e_0 = 0.02$ or $e_0 = 0.15$, respectively). The initial eccentricities of the planetesimal orbits in the protoplanetary disk were small. However, they could increase due to the mutual gravitational influence of the planetesimals. Ipatov (1993, 2000) noted that that the average eccentricity of the orbits of bodies in the feeding zone of the terrestrial planets could increase to 0.2 and even 0.4 in the final stages of planetesimal disk evolution. In the feeding zone of the giant planets, the increase in the orbital eccentricities of planetesimals was even greater than in the feeding zone of the terrestrial planets. In each calculation variant (with fixed $a_{min}$ and $e_0$) in the Proxima Centauri system for $(i + 1)$th planetesimal initial $a$ value was calculated by the formula $a_{0(i+1)} = (a_{0i}^2 + [a_{max}^2 - a_{min}^2]/N_0)^{1/2}$, where $a_{0i}$ is the $a_0$ value for the $i$th planetesimal, $a_{max} = a_{min} + 0.1$ AU, $N_0 = 250$. The values of $a_{min}$ varied from 0.9 to 2.2 AU with a step of 0.1 AU. The time interval in the calculations was at least 100 million years (if the evolution did not end sooner). In the variants with $a_{min}$ from 1.2 to 1.7 AU calculations were carried out for the evolution over several hundred million years (up to 1 billion years).

The equations of motion were integrated using the symplectic algorithm from the SWIFT package (Levison and Duncan, 1994). It was noted (Frantseva et al., 2022) that the integration step in this algorithm decreases significantly at distances smaller than 3.5 Hill radii. Planetesimals that collided with planets or the star or reached 1200 AU from the star were excluded from the integration. According to Schwarz et al. (2018), the radius of the Hill sphere of Proxima Centauri is 1200 AU. In the present paper, the primary focus is on the study of the motion of planetesimals in the outer part of the Hill sphere of Proxima Centauri, at distances of 500 to 1200 AU from the star.

Ejection of planetesimals from the feeding zone of planet $c$ to a distance $R$ greater than 500 AU from the star was briefly discussed in (Ipatov, 2022, 2023; Marov and Ipatov, 2023). In calculations over the time interval $T \geq 10$ Myr for present-day mass of planet $c$ ($k_c = 1$), the ratio $p_{cej} = p_c/p_{ej}$, where $p_c$ is the probability of collision of a planetesimal with planet $c$ and $p_{ej}$ is the probability of ejection of a planetesimal to a distance $R > 500$ AU, at $e_0 = 0.02$ and $e_0 = 0.15$ was in the ranges of 0.8–1.3 and 0.4–0.6, respectively. With a planet mass less than its present mass by a factor of two ($k_c = 0.5$) and $T \geq 100$ Myr, this ratio was in the ranges of 1.3–1.5 and 0.5–0.6. For a small mass of planet $c$, the ejection of planetesimals was minimal. The total mass of planetesimals from the feeding zone of planet $c$ that reached 500 AU from the star, could be about $(3.5–7)$ $m_E$, and the total mass of planetesimals in this zone could be at least $10m_E$ and $15m_E$ at $e_0 = 0.02$ and $e_0 = 0.15$, respectively. Based on the $p_{ej}$ obtained from the law of conservation of energy, it can be estimated that the semimajor axis of the orbit of planet $c$ could decrease by at least a factor of 1.5 during its accumulation.

## TIMESCALES BEFORE THE INCREASE OF PLANETESIMAL DISTANCES FROM THE STAR TO 500 AU

In Table 1, values of the fraction $p_{ej}$ of planetesimals that reached 500 AU from the star over the entire time interval (among all initial planetesimals) are provided for the present mass of planet $c$ and a range of $a_{min}$ values. In addition, it includes the ratios $f_{10}, f_{10-50}, f_{50-100}, f_{>100}$, which represent the number of planetesimals that reached 500 AU from the star in 10 million years, between 10 and 50 million years, between 50

**Table 1.** Probabilities $p_{ej}$ of ejection of planetesimals to a distance from the star greater than 500 AU and fractions $f_{10}$, $f_{10-50}$, $f_{50-100}$, $f_{>100}$ of planetesimals ejected over several time intervals at the present mass of planet $c$ ($k_c = 1$), $e_0 = 0.02$ or $e_0 = 0.15$, for several $a_{min}$ values from 1.0 to 2.2 AU given in the first row of the table

| $e_0$ | $a_{min}$, AU | 1.0 | 1.1 | 1.2 | 1.3 | 1.4 | 1.5 | 1.6 | 1.7 | 1.8 | 1.9 | 2.0 | 2.1 | 2.2 |
|---|---|---|---|---|---|---|---|---|---|---|---|---|---|---|
| 0.02 | $p_{ej}$ | | 0.03 | 0.34 | 0.46 | 0.48 | 0.48 | 0.44 | 0.57 | 0.12 | | | | |
| 0.15 | $p_{ej}$ | 0.17 | 0.60 | 0.63 | 0.65 | 0.73 | 0.71 | 0.72 | 0.62 | 0.50 | 0.42 | 0.95 | 0.87 | 0.31 |
| 0.02 | $f_{10}$ | | 0.71 | 0.68 | 0.96 | 0.77 | 0.87 | 0.96 | 0.96 | 0.39 | | | | |
| 0.02 | $f_{10-50}$ | | 0.19 | 0.27 | 0.04 | 0.14 | 0.09 | 0.03 | 0.03 | 0.50 | | | | |
| 0.02 | $f_{50-100}$ | | 0.1 | 0.04 | 0 | 0.03 | 0.02 | 0.01 | 0 | 0.11 | | | | |
| 0.02 | $f_{>100}$ | | – | 0.01 | – | 0.06 | 0.02 | – | 0.01 | – | | | | |
| 0.15 | $f_{10}$ | 0.69 | 0.90 | 0.91 | 0.95 | 0.95 | 0.96 | 0.97 | 0.93 | 0.94 | 0.85 | 0.82 | 0.41 | 0.12 |
| 0.15 | $f_{10-50}$ | 0.19 | 0.07 | 0.06 | 0.05 | 0.045 | 0.04 | 0.03 | 0.064 | 0.06 | 0.14 | 0.17 | 0.53 | 0.57 |
| 0.15 | $f_{50-100}$ | 0.12 | 0.03 | 0.01 | 0 | 0.005 | 0 | 0 | 0 | 0 | 0.01 | 0.01 | 0.06 | 0.31 |
| 0.15 | $f_{>100}$ | – | 0 | 0.02 | – | 0 | 0 | – | 0.006 | – | – | 0 | – | – |

$p_{ej}$ is the fraction of planetesimals reaching 500 AU from the star over the entire time interval (among all the initial planetesimals). $f_{10}$, $f_{10-50}$, $f_{50-100}$, $f_{>100}$ represent the ratios of the number of planetesimals that reached 500 AU from the star in 10 million years, between 10 and 50 million years, between 50 and 100 million years, and over 100 million years, respectively, divided by the number of planetesimals that reached 500 AU from the star over the entire time interval. A dash in the $f_{>100}$ column indicates that the time interval in this variant of calculations was 100 Myr. In each calculation variant, 250 bodies with orbital semimajor axes ranging from $a_{min}$ to $a_{min} + 0.1$ AU, eccentricities $e_0$ and inclinations $e_0/2$ rad were considered at the initial time.

and 100 million years, and over 100 million years, respectively, divided by the total number of planetesimals that have reached 500 AU from the star over the entire time interval. The rest of the planetesimals either collided with planets or remained in elliptical orbits with aphelion distances $Q < 500$ AU. The $p_{ej}$ values were on average about 0.5 for $e_0 = 0.02$ and $1.3 \le a_{min} \le 1.7$ AU, and about 0.6–0.7 for $e_0 = 0.15$ and $1.2 \le a_{min} \le 1.7$ AU. Large $p_{ej}$ values, equal to 0.95 and 0.87 for $e_0 = 0.15$ and $a_{min}$ of 2.0 and 2.1 AU were due to the fact that in these variants, the fraction of planetesimals that collided with the planets was small. For $a_{min}$, equal to 1.3, 1.6 and 1.7 AU, and $e_0 = 0.02$, $f_{10}$ was equal to 0.96. The $f_{10}$ values for $1.1 \le a_{min} \le 1.8$ AU. and $e_0 = 0.15$ were not less than 0.9. In these cases, almost all planetesimals reached 500 AU in less than 10 million years. The predominant ejection of planetesimals after 10 Myr occurred only for $2.1 \le a_{min} \le 2.2$ AU and $e_0 = 0.15$.

As noted above, the main ejection of planetesimals from the feeding zone of planet $c$ to distances $R > 500$ AU from the star occurred within the first 10 million years. The results of calculations for larger time intervals are presented for the current mass of planet $c$. Values of $f_{10-50}$ and $f_{50-100}$ are greater for planetesimals that are initially farther from the orbit of planet $c$. For $e_0 = 0.15$ and $a_{min} = 1.0$ AU or $a_{min} = 2.2$ AU, as well as for $e_0 = 0.02$ and $a_{min} = 1.8$ AU, $f_{50-100} > 0.1$ was obtained. In some calculation variants, planetesimals reached 500 AU over time even longer than 100 million years, but the $f_{>100}$ values in Table 1 are small. For example, the fraction $f_{100-500}$ of planetesimals that reached 500 AU over $100 < t < 500$ Myr (among all ejected planetesimals) was $3/160 \approx 0.02$ for $a_{min} = 1.2$ AU and $e_0 = 0.15$. In the case of $e_0 = 0.02$ for $a_{min} = 1.4$ AU and $a_{min} = 1.7$ AU, $f_{100-500}$ was $7/120 \approx 0.058$ and $2/143 \approx 0.014$, respectively, and $f_{100-500} > f_{50-100}$. Some planetesimals (for example, at $a_{min} = 1.2$ AU and $e_0 = 0.15$) were also ejected after 500 Myr.

With a smaller mass of the planet moving in the orbit of planet $c$, planetesimals reached 500 AU later than with the present mass of planet $c$ (compare Tables 1 and 2). For $a_{min} = 1.4$ and $a_{min} = 1.5$ AU, when the ratio $k_c$ of the mass of the planet to the present mass of planet $c$ was 0.5, the $f_{10-50}$ values were several times greater than for $k_c = 1$. However, the $f_{10}$ values in these variants were not small and amounted to about 0.6 and 0.5 at $e_0 = 0.15$ and $e_0 = 0.02$, respectively (see Table 2). Therefore, in the last doubling of the mass of the growing planet $c$, more than half of the ejected planetesimals were ejected within the first 10 million years. For a planetesimal with a mass that was 10 times less than the present mass of planet $c$, in the calculation variants presented in Table 2, the fraction $p_{ej}$ of planetesimals that reached distances greater than 500 AU from the star was less than 0.2, and $f_{10}$ was close to zero, i.e., there was almost no ejection of planetesimals outside the Hill sphere in the first 10 Myr. For the Proxima Centauri system, over a period of several million years, the mass of the growing planet could exceed half the present mass of planet $c$. The main ejection of planetesimals from this system occurred after planet $c$ reached approximately half of its mass. Estimates with small values of the planet's mass may be of interest for

**Table 2.** Probabilities $p_{ej}$ of ejection of planetesimals to distances from the star greater than 500 AU and fractions $f_{10}$, $f_{10-50}$, $f_{50-100}$ of planetesimals ejected over several time intervals at a planet mass equal to $k_c m_c$ ($m_c$ is the present mass of planet $c$, $k_c = 0.5$ or $k_c = 0.1$), $e_0 = 0.02$ or $e_0 = 0.15$, and $a_{min} = 1.4$ or $a_{min} = 1.5$ AU

| $k_c$ | 0.5 | 0.5 | 0.5 | 0.5 | 0.1 | 0.1 | 0.1 | 0.1 |
|---|---|---|---|---|---|---|---|---|
| $E_0$ | 0.02 | 0.02 | 0.15 | 0.15 | 0.02 | 0.02 | 0.15 | 0.15 |
| $a_{min}$, AU | 1.4 | 1.5 | 1.4 | 1.5 | 1.4 | 1.5 | 1.4 | 1.5 |
| $p_{ej}$ | 0.284 | 0.344 | 0.508 | 0.664 | 0.036 | 0.068 | 0.14 | 0.188 |
| $f_{10}$ | 0.47 | 0.49 | 0.63 | 0.60 | 0 | 0 | 0.03 | 0 |
| $f_{10-50}$ | 0.52 | 0.45 | 0.31 | 0.37 | 0.44 | 0.18 | 0.37 | 0.47 |
| $f_{50-100}$ | 0.01 | 0.06 | 0.06 | 0.03 | 0.56 | 0.82 | 0.60 | 0.53 |

$p_{ej}$ is the fraction of planetesimals reaching 500 AU from the star over the entire time interval $T = 100$ million years. $f_{10}$, $f_{10-50}$, $f_{50-100}$ represent the ratio of the number of planetesimals that reached 500 AU from the star in the first 10 million years, between 10 and 50 million years, and between 50 and 100 million years, to the number of planetesimals that reached 500 AU from the star over $T = 100$ million years. In each calculation variant, 250 bodies with orbital semimajor axes from $a_{min}$ to $a_{min} + 0.1$ AU, eccentricities $e_0$ and inclinations $e_0/2$ rad were considered at the initial time.

other possible exoplanetary systems. It should be noted that for other planetary systems, the times at which planetesimals reach 500 AU from the star depend not only on the masses of the star and the planet, but also on the distance from the star to the planet. The farther planet is from the star, the longer the time between close encounters of planetesimals with the planet and the longer the characteristic time to reach 500 AU from the star.

## ECCENTRICITIES AND ORBITAL INCLINATIONS OF PLANETESIMALS AT 500 AND 1200 AU FROM THE STAR

Table 3 shows the orbital eccentricities of planetesimals at the times when their distances from the star first reached 500 AU. $N_{<4}$, $N_{4-6}$, $N_{6-8}$, $N_{8-10}$, $N_{002}$, $N_{005}$, $N_{01}$, $N_{11}$, and $N_{>11}$ denote the number of bodies whose orbital eccentricities fall within the intervals ($e_n$, $e_x$) from 0.994 to 0.996, from 0.996 to 0.998, from 0.998 to 1.0, from 1.0 to 1.002, from 1.002 to 1.005, from 1.005 to 1.01, from 1.01 to 1.1, and greater than 1.1, respectively. $N_s$ is the total number of planetesimals that first reached 500 AU from the star. $e_{min}$ and $e_{max}$ are the minimum and maximum eccentricities of the orbits. $\Sigma_n$ is the sum of the numbers of planetesimals in the column for the variants presented above. $\Sigma_f = \Sigma_n/N_s$ represents the fractions of planetesimals with different eccentricities. The mass $m_p$ of the planet moving in the orbit of planet $c$ equaled $7k_c m_E$. In most calculations, $m_p = 7m_E$. Values of the planet's mass equal to $3.5m_E$ ($k_c = 0.5$) or $0.7m_E$ ($k_c = 0.1$) were also considered. In most calculation variants, the integration step $t_s$ was equal to one Earth day. The calculation results for $t_s$ equal to 0.2, 0.5, or two Earth days are marked in the tables with *, **, or ***, respectively. Calculations with different integration steps $t_s$ yielded approximately the same results (considering that the evolution of orbits becomes chaotic during close encounters). Table 4 shows the orbital eccentricities of planetesimals at the times when their distances from the star first reached 1200 AU. The designations are the same as in Table 3. Although the eccentricities of the orbits of the planetesimals moving at distances greater than 500 AU from the star are quite large, these orbits differed from the orbits of exocomets considered in (Schwarz et al., 2018). The perihelions of such planetesimals often came close to the orbit of planet $c$, while Schwarz et al. (2018) considered initial orbits with perihelion distances less than 0.0485 AU.

With the current mass of planet $c$ among planetesimals reaching 500 AU from the star, the fraction of planetesimals with orbital eccentricities greater than 1 was 0.05 and 0.1 for the initial eccentricities of their orbits $e_0 = 0.02$ and $e_0 = 0.15$, respectively. Among the planetesimals that reached 1200 AU from the star, this fraction was about 0.3 for both $e_0$ values. The minimum values of the orbital eccentricity of planetesimals that reached 500 and 1200 AU from the star were 0.992 and 0.995, respectively.

When considering the present mass of planet $c$ ($k_c = 1$), orbital eccentricities $e_{500}$ of the planetesimals that reached $a_{lim} = 500$ AU from the star were in the range from 0.992 to 0.998 in 87 and 82% of cases for $e_0 = 0.02$ and $e_0 = 0.15$, respectively (Table 3). In all considered cases, they exceeded 0.992 and were less than 1.4. The fraction of planetesimals with $e_{500} > 1$ was 0.05 and 0.1 for $e_0 = 0.02$ and $e_0 = 0.15$, respectively (Table 3). There was no significant dependence on $a_{min}$ for the distribution of planetesimals by eccentricity $e_{500}$. However, a slight dependence may be observed. For example, for $e_0 = 0.15$ and $k_c = 1$, the fraction of orbits with eccentricities $e < 0.994$ at $a_{lim} = 500$ AU and with $0.996 < e < 0.998$ at $a_{lim} = 1200$ AU was slightly higher for $a_{min} > 1.8$ AU than the values averaged over all $a_{min}$ values.

**Table 3.** Distribution of planetesimals that first reached 500 AU from the star, over the intervals of eccentricities ($e_n$, $e_x$) of their orbits for $e_0$ equal to 0.02 or 0.15, and several $a_{min}$ values from 1.1 to 2.2 AU given in the first column of the table

| $a_{min}$, AU | $e_0$ | $k_c$ | $N_{<4}$ | $N_{4-6}$ | $N_{6-8}$ | $N_{8-10}$ | $N_{002}$ | $N_{005}$ | $N_{01}$ | $N_{11}$ | $N_{>11}$ | $N_s$ | $e_{min}$ | $e_{max}$ |
|---|---|---|---|---|---|---|---|---|---|---|---|---|---|---|
| $e_n=$ | | | — | 0.994 | 0.996 | 0.998 | 1.0 | 1.002 | 1.005 | 1.01 | 1.1 | | | |
| $e_x=$ | | | 0.994 | 0.996 | 0.998 | 1.0 | 1.002 | 1.005 | 1.01 | 1.1 | — | | | |
| 1.1 | 0.02 | 1 | 2 | 3 | 2 | 0 | 1 | 0 | 0 | 0 | 0 | 8 | 0.993 | 1.000 |
| 1.2 | 0.02 | 1 | 29 | 25 | 17 | 9 | 1 | 2 | 0 | 2 | 0 | 85 | 0.992 | 1.021 |
| 1.3 | 0.02 | 1 | 29 | 48 | 26 | 8 | 2 | 1 | 0 | 0 | 0 | 114 | 0.993 | 1.002 |
| 1.4 | 0.02 | 1 | 29 | 45 | 25 | 8 | 3 | 5 | 2 | 2 | 1 | 120 | 0.992 | 1.173 |
| 1.4 | 0.02 | 1* | 23 | 40 | 15 | 9 | 2 | 1 | 1 | 1 | 0 | 92 | 0.992 | 1.020 |
| 1.4 | 0.02 | 1** | 26 | 36 | 17 | 6 | 3 | 2 | 0 | 4 | 0 | 94 | 0.993 | 1.073 |
| 1.4 | 0.02 | 1*** | 18 | 35 | 22 | 5 | 7 | 1 | 0 | 1 | 0 | 89 | 0.992 | 1.031 |
| 1.5 | 0.02 | 1 | 31 | 56 | 21 | 11 | 0 | 0 | 0 | 0 | 0 | 119 | 0.992 | 1.000 |
| 1.6 | 0.02 | 1 | 40 | 37 | 19 | 10 | 2 | 1 | 0 | 0 | 0 | 109 | 0.992 | 1.003 |
| 1.7 | 0.02 | 1 | 42 | 62 | 27 | 10 | 2 | 0 | 0 | 0 | 0 | 143 | 0.992 | 1.001 |
| 1.8 | 0.02 | 1 | 11 | 13 | 6 | 1 | 0 | 0 | 0 | 0 | 0 | 31 | 0.993 | 0.999 |
| $\Sigma_n$ | 0.02 | 1 | 280 | 400 | 197 | 77 | 23 | 13 | 3 | 10 | 1 | 1004 | | |
| $\Sigma_f$ | 0.02 | 1 | 0.279 | 0.399 | 0.196 | 0.077 | 0.023 | 0.013 | 0.003 | 0.010 | 0.001 | 1.00 | 0.992 | 1.173 |
| 1.0 | 0.15 | 1 | 11 | 18 | 7 | 3 | 1 | 1 | 1 | 0 | 0 | 42 | 0.993 | 1.007 |
| 1.1 | 0.15 | 1 | 26 | 37 | 14 | 8 | 6 | 4 | 1 | 1 | 1 | 98 | 0.992 | 1.176 |
| 1.2 | 0.15 | 1 | 32 | 66 | 29 | 10 | 8 | 4 | 5 | 5 | 1 | 160 | 0.993 | 1.167 |
| 1.3 | 0.15 | 1 | 29 | 56 | 34 | 26 | 7 | 1 | 1 | 3 | 1 | 158 | 0.992 | 1.135 |
| 1.3 | 0.15 | 1** | 28 | 65 | 28 | 21 | 10 | 4 | 3 | 2 | 1 | 162 | 0.993 | 1.125 |
| 1.4 | 0.15 | 1 | 37 | 68 | 34 | 19 | 8 | 5 | 3 | 8 | 1 | 183 | 0.992 | 1.132 |
| 1.4 | 0.15 | 1* | 38 | 67 | 37 | 19 | 6 | 4 | 1 | 2 | 2 | 176 | 0.992 | 1.356 |
| 1.4 | 0.15 | 1** | 29 | 69 | 41 | 14 | 14 | 8 | 1 | 5 | 1 | 182 | 0.993 | 1.164 |
| 1.4 | 0.15 | 1*** | 33 | 67 | 31 | 20 | 8 | 4 | 2 | 9 | 0 | 174 | 0.993 | 1.064 |
| 1.5 | 0.15 | 1 | 32 | 68 | 38 | 14 | 12 | 4 | 2 | 6 | 2 | 178 | 0.993 | 1.235 |
| 1.5 | 0.15 | 1** | 33 | 73 | 28 | 20 | 7 | 4 | 0 | 11 | 1 | 177 | 0.993 | 1.129 |
| 1.5 | 0.15 | 1*** | 27 | 67 | 29 | 20 | 11 | 7 | 2 | 5 | 4 | 172 | 0.993 | 1.294 |
| 1.6 | 0.15 | 1 | 34 | 66 | 34 | 17 | 11 | 7 | 6 | 3 | 2 | 180 | 0.993 | 1.180 |
| 1.7 | 0.15 | 1 | 37 | 59 | 27 | 12 | 6 | 4 | 3 | 7 | 1 | 156 | 0.992 | 1.191 |
| 1.8 | 0.15 | 1 | 31 | 58 | 29 | 4 | 0 | 0 | 1 | 1 | 1 | 125 | 0.992 | 1.189 |
| 1.9 | 0.15 | 1 | 28 | 51 | 17 | 7 | 0 | 1 | 1 | 0 | 0 | 105 | 0.992 | 1.005 |
| 2.0 | 0.15 | 1 | 79 | 99 | 46 | 12 | 2 | 0 | 0 | 0 | 0 | 238 | 0.992 | 1.001 |
| 2.1 | 0.15 | 1 | 74 | 96 | 35 | 9 | 1 | 0 | 1 | 0 | 1 | 217 | 0.992 | 1.169 |
| 2.2 | 0.15 | 1 | 25 | 38 | 9 | 6 | 0 | 0 | 0 | 0 | 0 | 78 | 0.992 | 0.999 |
| $\Sigma_n$ | 0.15 | 1 | 652 | 1200 | 540 | 257 | 117 | 57 | 33 | 68 | 20 | 2919 | | |
| $\Sigma_f$ | 0.15 | 1 | 0.224 | 0.411 | 0.185 | 0.088 | 0.040 | 0.020 | 0.011 | 0.023 | 0.007 | 1.0 | 0.992 | 1.356 |
| 1.4 | 0.02 | 0.5 | 23 | 33 | 7 | 2 | 2 | 0 | 1 | 3 | 0 | 71 | 0.993 | 1.066 |
| 1.5 | 0.02 | 0.5 | 30 | 38 | 12 | 2 | 0 | 0 | 1 | 2 | 1 | 86 | 0.993 | 1.264 |
| $\Sigma_n$ | 0.02 | 0.5 | 53 | 71 | 19 | 4 | 2 | 0 | 2 | 5 | 1 | 157 | | |
| $\Sigma_f$ | 0.02 | 0.5 | 0.338 | 0.452 | 0.121 | 0.025 | 0.013 | 0 | 0.013 | 0.032 | 0.006 | 0.000 | 0.993 | 1.264 |
| 1.4 | 0.15 | 0.5 | 30 | 55 | 22 | 6 | 5 | 1 | 1 | 6 | 1 | 127 | 0.993 | 1.108 |

**Table 3.** (Contd.)

| $a_{min}$, AU | $e_0$ | $k_c$ | $N_{<4}$ | $N_{4-6}$ | $N_{6-8}$ | $N_{8-10}$ | $N_{002}$ | $N_{005}$ | $N_{01}$ | $N_{11}$ | $N_{>11}$ | $N_s$ | $e_{min}$ | $e_{max}$ |
|---|---|---|---|---|---|---|---|---|---|---|---|---|---|---|
| 1.5 | 0.15 | 0.5 | 40 | 83 | 24 | 4 | 6 | 3 | 1 | 4 | 1 | 166 | 0.993 | 1.105 |
| $\Sigma_n$ | 0.15 | 0.5 | 70 | 138 | 46 | 10 | 11 | 4 | 2 | 10 | 2 | 293 | | |
| $\Sigma_f$ | 0.15 | 0.5 | 0.239 | 0.471 | 0.157 | 0.034 | 0.038 | 0.014 | 0.007 | 0.034 | 0.007 | 1.0 | 0.993 | 1.108 |
| 1.4 | 0.02 | 0.1 | 2 | 3 | 2 | 1 | 1 | 0 | 0 | 0 | 0 | 9 | 0.994 | 1.002 |
| 1.5 | 0.02 | 0.1 | 1 | 8 | 3 | 2 | 1 | 0 | 0 | 2 | 0 | 17 | 0.994 | 1.056 |
| $\Sigma_n$ | 0.02 | 0.1 | 3 | 11 | 5 | 3 | 2 | 0 | 0 | 2 | 0 | 26 | | |
| $\Sigma_f$ | 0.02 | 0.1 | 0.115 | 0.423 | 0.192 | 0.115 | 0.077 | 0 | 0 | 0.077 | 0 | 1.0 | 0.994 | 1.056 |
| 1.4 | 0.15 | 0.1 | 9 | 16 | 2 | 1 | 2 | 2 | 1 | 2 | 0 | 35 | 0.993 | 1.018 |
| 1.5 | 0.15 | 0.1 | 19 | 24 | 4 | 2 | 3 | 1 | 1 | 1 | 0 | 55 | 0.993 | 1.055 |
| $\Sigma_n$ | 0.15 | 0.1 | 28 | 40 | 6 | 3 | 5 | 3 | 2 | 3 | 0 | 90 | | |
| $\Sigma_f$ | 0.15 | 0.1 | 0.311 | 0.444 | 0.067 | 0.033 | 0.056 | 0.033 | 0.022 | 0.033 | 0 | 1.0 | 0.993 | 1.055 |

$N_{<4}$, $N_{4-6}$, $N_{8-10}$, $N_{002}$, $N_{005}$, $N_{01}$, $N_{11}$, $N_{>11}$ represent the number of planetesimals whose orbital eccentricities fall within the intervals $(e_n, e_x)$ from 0.994 to 0.996, from 0.996 to 0.998, from 0.998 to 1.0, from 1.0 to 1.002, from 1.002 to 1.005, from 1.005 to 1.01, from 1.01 to 1.1, and greater than 1.1, respectively. $N_s$ is the total number of planetesimals that reached 500 AU from the star. $e_{min}$ and $e_{max}$ are the minimum and maximum eccentricities of the orbits. In each calculation variant, 250 planetesimals with orbital semimajor axes from $a_{min}$ to $a_{min} + 0.1$ AU, eccentricities $e_0$ and inclinations $e_0/2$ rad, respectively, were considered at the initial time. The mass of planet $c$ was 7 $k_c m_E$. $\Sigma_n$ is the sum of the numbers of planetesimals in the column for the variants presented above. $S_f = S_n/N_s$.

Similar (to the case of $k_c = 1$) $e_{500}$ values were obtained for the embryo of planet $c$ with a mass equal to $3.5m_E$ ($k_c = 0.5$): $0.992 \le e_{500} \le 0.998$ in 91 and 87% of cases for $e_0 = 0.02$ and $e_0 = 0.15$, respectively, and the fraction of planetesimals with $e_{500} > 1$ was 6 and 10% for $e_0 = 0.02$ and $e_0 = 0.15$, respectively (Table 3). At $k_c = 0.1$, there were slightly more planetesimals with $e_{500} > 1$: 12.5 and 14.4% for $e_0 = 0.02$ and $e_0 = 0.15$, respectively (Table 3). These estimates indicate that the distribution of planetesimals that reached 500 AU from the star in terms of their eccentricities did not differ significantly at different times of planet $c$ accumulation (for different considered values of the mass of the planet moving in the orbit of planet $c$).

Orbital eccentricities $e_{1200}$ of planetesimals that reached $a_{lim} = 1200$ AU from the star were on average slightly larger than $e_{500}$. In all variants presented in Table 4, $e_{1200} \ge 0.995$, and the fraction of planetesimals with $e_{1200} < 0.996$ did not exceed 0.01. For all $k_c$ and $e_0$ values in Table 4, the fraction $\Sigma_f$ of the number of bodies with eccentricities $0.996 \le e_{1200} \le 1$ was not less than 0.6. Among all planetesimals that reached 1200 AU, at $k_c = 1$, the fraction of orbits with eccentricities greater than 1 was approximately 0.306 and 0.315 for $e_0 = 0.02$ and $e_0 = 0.15$, respectively. At $k_c = 1$, the fraction of planetesimals with $e_{1200} > 1.1$ did not exceed 0.01. In calculations with $k_c = 0.1$, this fraction was equal to 0 for $a_{lim} = 500$ AU and was close to 0.1 for $a_{lim} = 1200$ AU.

For comparison, Ipatov (1989b) found that during the evolution of a disk of bodies with a total mass equal to $200m_E$, which corresponds to the feeding zone of Uranus and Neptune, 13% of the orbits of ejected bodies had $e > 1.1$. Ipatov (1989a, 1989b) considered the migration of bodies from the feeding zones of Uranus and Neptune. The gravitational influence of nearly formed planets in the Solar System and bodies in the disk was taken into account using the method of spheres of action. In (Ipatov, 1989a), the total mass of disk bodies was $10m_E$, and the masses of the bodies were an order of magnitude smaller than in (Ipatov, 1989b). The distribution of bodies ejected into hyperbolic orbits according to their eccentricities and the Tisserand parameter was considered. It was found that 30% of the ejected bodies had $e > 1.1$. Although the initial bodies were in the feeding zone of Uranus and Neptune, the values of the Tisserand parameter indicate that Jupiter ejected more bodies than Uranus or Neptune. Wherein, Jupiter ejected bodies into more eccentric orbits.

Tables 5 and 6 show the orbital inclinations of planetesimals that first reached distances of 500 and 1200 AU from the star, respectively. $N_2$, $N_4$, $N_6$, $N_8$, $N_{10}$, $N_{15}$, $N_{20}$, $N_{>20}$ represent the number of planetesimals whose orbital inclinations fall within the intervals $(i_n, i_x)$ of 0° to 2°, 2° to 4°, 4° to 6°, 6° to 8°, 8° to 10°, 10° to 15°, 15° to 20°, and greater than 20°, respectively.

Orbital inclinations $i_{500}$ and $i_{1200}$ of more than 80% of planetesimals that reached 500 or 1200 AU, respec-

**Table 4.** Distribution of planetesimals that first reached 1200 AU from the star by intervals of their orbital eccentricities ($e_n$, $e_x$) for $e_0$ equal to 0.02 or 0.15 and several $a_{min}$ values from 1.1 to 2.2 AU given in the first column of the table

| $a_{min}$, AU | $e_0$ | $k_c$ | $N_{4-6}$ | $N_{6-8}$ | $N_{8-10}$ | $N_{002}$ | $N_{005}$ | $N_{01}$ | $N_{11}$ | $N_{>11}$ | $N_s$ | $e_{min}$ | $e_{max}$ |
|---|---|---|---|---|---|---|---|---|---|---|---|---|---|
| $e_n=$ | | | 0.994 | 0.996 | 0.998 | 1.0 | 1.002 | 1.005 | 1.01 | 1.1 | | | |
| $e_x=$ | | | 0.996 | 0.998 | 1.0 | 1.002 | 1.005 | 1.01 | 1.1 | — | | | |
| 1.2 | 0.02 | 1 | 0 | 27 | 29 | 11 | 13 | 2 | 3 | 0 | 85 | 0.996 | 1.083 |
| 1.3 | 0.02 | 1 | 0 | 27 | 55 | 14 | 9 | 7 | 2 | 0 | 114 | 0.996 | 1.024 |
| 1.4 | 0.02 | 1 | 0 | 32 | 34 | 21 | 15 | 8 | 6 | 1 | 117 | 0.997 | 1.136 |
| 1.5 | 0.02 | 1 | 1 | 40 | 46 | 14 | 10 | 4 | 4 | 0 | 119 | 0.996 | 1.023 |
| 1.6 | 0.02 | 1 | 1 | 40 | 37 | 15 | 11 | 3 | 1 | 1 | 109 | 0.996 | 1.242 |
| 1.7 | 0.02 | 1 | 5 | 59 | 45 | 14 | 9 | 4 | 7 | 0 | 143 | 0.995 | 1.017 |
| $\Sigma_n$ | 0.02 | 1 | 7 | 225 | 246 | 89 | 67 | 28 | 23 | 2 | 687 | | |
| $\Sigma_f$ | 0.02 | 1 | 0.010 | 0.328 | 0.358 | 0.130 | 0.098 | 0.041 | 0.033 | 0.003 | 1.0 | 0.995 | 1.242 |
| 1.1 | 0.15 | 1 | 1 | 28 | 36 | 12 | 11 | 1 | 6 | 1 | 96 | 0.996 | 1.242 |
| 1.2 | 0.15 | 1 | 0 | 49 | 50 | 29 | 15 | 8 | 8 | 1 | 160 | 0.996 | 1.151 |
| 1.3 | 0.15 | 1 | 0 | 45 | 63 | 18 | 17 | 9 | 5 | 1 | 158 | 0.996 | 1.105 |
| 1.4 | 0.15 | 1 | 0 | 51 | 68 | 29 | 15 | 10 | 10 | 0 | 183 | 0.996 | 1.096 |
| 1.5 | 0.15 | 1 | 1 | 63 | 56 | 25 | 15 | 7 | 9 | 2 | 178 | 0.996 | 1.240 |
| 1.5 | 0.15 | 1** | 0 | 43 | 77 | 19 | 15 | 12 | 10 | 1 | 177 | 0.996 | 1.127 |
| 1.5 | 0.15 | 1*** | 1 | 44 | 59 | 31 | 20 | 5 | 8 | 4 | 172 | 0.996 | 1.309 |
| 1.6 | 0.15 | 1 | 0 | 51 | 58 | 30 | 28 | 5 | 6 | 2 | 180 | 0.996 | 1.173 |
| 1.7 | 0.15 | 1 | 1 | 49 | 51 | 22 | 16 | 5 | 11 | 1 | 156 | 0.995 | 1.199 |
| 1.8 | 0.15 | 1 | 1 | 54 | 38 | 11 | 9 | 7 | 3 | 1 | 124 | 0.996 | 1.249 |
| 1.9 | 0.15 | 1 | 0 | 38 | 39 | 14 | 12 | 2 | 0 | 0 | 105 | 0.996 | 1.007 |
| 2.0 | 0.15 | 1 | 1 | 89 | 87 | 24 | 21 | 15 | 0 | 0 | 237 | 0.996 | 1.010 |
| 2.1 | 0.15 | 1 | 1 | 87 | 85 | 24 | 14 | 6 | 1 | 1 | 219 | 0.996 | 1.157 |
| 2.2 | 0.15 | 1 | 1 | 32 | 31 | 10 | 4 | 5 | 1 | 0 | 84 | 0.996 | 1.018 |
| $\Sigma_n$ | 0.15 | 1 | 8 | 723 | 798 | 298 | 212 | 97 | 78 | 15 | 2229 | | |
| $\Sigma_f$ | 0.15 | 1 | 0.003 | 0.326 | 0.357 | 0.134 | 0.094 | 0.045 | 0.034 | 0.007 | 1.0 | 0.995 | 1.309 |
| 1.4 | 0.02 | 0.5 | 0 | 34 | 25 | 4 | 2 | 1 | 5 | 0 | 71 | 0.997 | 1.063 |
| 1.5 | 0.02 | 0.5 | 0 | 36 | 35 | 4 | 5 | 1 | 4 | 1 | 86 | 0.996 | 1.173 |
| $\Sigma_n$ | 0.02 | 0.5 | 0 | 70 | 60 | 8 | 7 | 2 | 9 | 1 | 157 | | |
| $\Sigma_f$ | 0.02 | 0.5 | 0 | 0.446 | 0.382 | 0.051 | 0.045 | 0.013 | 0.057 | 0.006 | 1.0 | 0.996 | 1.173 |
| 1.4 | 0.15 | 0.5 | 0 | 52 | 46 | 12 | 2 | 4 | 11 | 1 | 128 | 0.997 | 1.102 |
| 1.5 | 0.15 | 0.5 | 0 | 51 | 64 | 22 | 18 | 3 | 6 | 1 | 165 | 0.996 | 1.245 |
| $\Sigma_n$ | 0.15 | 0.5 | 0 | 103 | 110 | 34 | 20 | 7 | 17 | 2 | 293 | | |
| $\Sigma_f$ | 0.15 | 0.5 | 0 | 0.352 | 0.375 | 0.116 | 0.068 | 0.024 | 0.058 | 0.007 | 1.0 | 0.996 | 1.245 |
| 1.4 | 0.02 | 0.1 | 0 | 3 | 4 | 1 | 0 | 0 | 0 | 1 | 9 | 0.997 | 2.154 |
| $\Sigma_f$ | 0.02 | 0.1 | 0 | 0.33 | 0.44 | 0.11 | 0 | 0 | 0 | 0.11 | 1.0 | 0.997 | 2.154 |
| 1.4 | 0.15 | 0.1 | 0 | 11 | 10 | 4 | 4 | 2 | 1 | 3 | 35 | 0.997 | 1.581 |
| $\Sigma_f$ | 0.15 | 0.1 | 0 | 0.314 | 0.286 | 0.114 | 0.114 | 0.057 | 0.029 | 0.086 | 1.0 | 0.997 | 1.581 |

$N_{4-6}$, $N_{8-10}$, $N_{002}$, $N_{005}$, $N_{01}$, $N_{11}$, $N_{>11}$ represent the number of planetesimals whose orbital eccentricities fall within the intervals ($e_n$, $e_x$) from 0.994 to 0.996, from 0.996 to 0.998, from 0.998 to 1.0, from 1.0 to 1.002, from 1.002 to 1.005, from 1.005 to 1.01, from 1.01 to 1.1, and greater than 1.1, respectively. $N_s$ is the total number of planetesimals that reached 1200 AU from the star. $e_{min}$ and $e_{max}$ are the minimum and maximum orbital eccentricities. In each calculation variant, 250 planetesimals with orbital semimajor axes from $a_{min}$ to $a_{min}$ + 0.1 AU, eccentricities $e_0$ and inclinations $e_0/2$ rad, respectively, were considered at the initial time. The mass of the planet in the orbit of planet $c$ was $7k_c m_E$. $\Sigma_n$ is the sum of the numbers of planetesimals in the column for the cases considered. $\mathrm{S}_f = \mathrm{S}_n/\mathrm{N}_s$. The number $N_{<4}$ of planetesimals with orbital eccentricities less than 0.994 is equal to zero in all variants of Table 4.

tively, from the star did not exceed 10° (Tables 5–7). The $i_{500}$ and $i_{1200}$ values were generally close to each other, with significant differences between $i_{500}$ and $i_{1200}$ observed only at $k_c = 0.1$ and $e_0 = 0.02$ (Table 7). The fraction of planetesimals with $i > 20°$ in most series of calculations did not exceed 0.02 and, on average, for all variants with $k_c = 1$ did not exceed 0.01. Only when $k_c = 0.1$ and $e_0 = 0.15$, this fraction reached 0.07 and 0.11 for $a_{lim} = 500$ and $a_{lim} = 1200$ AU, respectively. The distribution of planetesimals' orbits by $i$ may depend on $a_{min}$. For example, in the case of $e_0 = 0.02$ and $k_c = 1$ (both for $a_{lim} = 500$ AU and $a_{lim} = 1200$ AU). the fraction of planetesimals with $i < 2°$ was lower for $a_{min} \leq 1.6$ AU and higher for $a_{min} \geq 1.7$ AU compared to the average value over all $a_{min}$ values. In the case of $e_0 = 0.15$ and $k_c = 1$ (both for $a_{lim} = 500$ AU and $a_{lim} = 1200$ AU), the fraction of planetesimals with $2° < i < 4°$ was more than 1.5 times higher for $a_{min} \geq 2$ AU, exceeding the average value over all $a_{min}$ values. For the two considered planetesimals at $a_{lim} = 1200$ AU, the orbital inclinations were 179° and 175°. However, even in these cases the orbits lay almost in the same plane as in the case of $i = 0$. For all other planetesimals, $i < 49°$ At $k_c = 1$, $a_{min} = 1.8$ AU, and $e_0 = 0.15$, the next lowest value of inclination $i_{200}$ less than 179° was 17.7°.

The calculation results presented in Tables 3–6 show relatively small differences in the distributions of planetesimals that reached 500 and 1200 AU over the eccentricities and inclinations of their orbits for various $k_c$ values (the mass of the planet in the orbit of planet $c$) and initial eccentricities $e_0$. Therefore, it can be expected that for more complex models of the planetesimal disk evolution during planet $c$ formation, these distributions of planetesimals will not differ significantly from the distributions given in Tables 3–6. The resulting distributions of planetesimals by orbital eccentricities and inclinations can be used as initial data for constructing models that take into account gravitational interactions and collisions of planetesimals and the influence of stars, simulating the formation of analogs of Hills and Oort clouds.

## ESTIMATES OF THE POSSIBILITY OF THE FORMATION OF HILLS AND OORT CLOUD ANALOGS AROUND PROXIMA CENTAURI

In Table 1, planetesimals were considered ejected when they reached 500 AU from the star. After completing these calculations, when no planetesimals remained in elliptical orbits or when the considered time interval reached several hundred million years (at least 100 million years), the time d$t$ during which the planetesimals increased their maximum distances from the star from 500 to 1200 AU was calculated. The fractions of planetesimals corresponding to certain d$t$ intervals for disks with different $a_{min}$ and $e_0$ values are presented in Tables 8–10. Most planetesimals (>84% at $e_0 = 0.02$ and >89% at $e_0 = 0.15$) increased their maximum distances from the star from 500 to 1200 AU in time d$t$ less than 1 million years. In Tables 8–9, the fraction of planetesimals with d$t < 0.1$ Myr was in the range of 0.32–0.53 for $e_0 = 0.02$ and 0.43–0.58 for $e_0 = 0.15$, i.e., on average, this fraction was higher for larger $e_0$. The fraction of planetesimals that increased their maximum distances from the star from 500 to 1200 AU in time d$t < 0.01$ Myr, was about 0.05–0.2 and 0.2–0.3 for $e_0 = 0.02$ and $e_0 = 0.15$, respectively. For d$t > 2$ Myr, this fraction was in the range of 0.05–0.08 and 0.01–0.06 for $e_0 = 0.02$ and $e_0 = 0.15$, respectively. The maximum d$t$ value, denoted as d$t_{max}$, in the considered calculation variants with 250 planetesimals ranged from 3 to 60 Myr. Most of the calculations presented in Tables 8–10 were performed with a time integration step $t_s = 1$ day. For $a_{min} = 1.5$ AU and $e_0 = 0.15$, Table 9 also shows the results of calculations with $t_s$ equal to 0.5 and 2 days. The difference in the results presented in Table 9 and obtained with different $t_s$ values is approximately the same as the difference between the results obtained for different $a_{min}$ values with the same $t_s$. Therefore, calculations with $t_s = 1$ day can be used to study the time during which the maximum distances of planetesimals from the star increased from 500 to 1200 AU.

Table 10 shows the fractions of planetesimals that increased their maximum distances from the star from 500 to 1200 AU for the present mass of planet $c$ ($k_c = 1$), as well as for the case when the ratio $k_c$ of the mass of the smaller planet to the mass of planet $c$ equals 0.5 (the orbits of both planets were the same). At $k_c = 0.5$ characteristic time d$t$ for planetesimals to increase their maximum distance from the star from 500 to 1200 AU was longer than at $k_c = 1$. In particular, for $a_{min}$ equal to 1.4 and 1.5 AU, the fraction of planetesimals with d$t < 0.1$ Myr was 0.23–0.28 at $k_c = 0.5$ and 0.46–0.52 at $k_c = 1$.

A large part (often up to 90%) of planetesimals that reached 500 AU from the star at $k_c = 1$ reached this distance in less than 10 million years. Most of the ejection of planetesimals after 10 Myr occurred only in the calculations with $a_{min} = 1.8$ AU and $e_0 = 0.02$, as well as at $e_0 = 0.15$ and $a_{min} = 2.1$ AU or $a_{min} = 2.2$ AU, i.e., for planetesimals from the outer part of the planet's feeding zone. For the present mass of the planet Proxima Centauri $c$, about 90% of the planetesimals that increased their maximum distances from the star from 500 to 1200 AU did so in less than 1 million years. With a total mass of matter ejected over 10 million years equal to $5m_E$ and a characteristic time for increasing the maximum distance of planetesimals from the star

**Table 5.** Distribution of planetesimals that first reached 500 AU from the star by inclination intervals $(i_n, i_x)$ of their orbits for $e_0$ equal to 0.02 or 0.15, and several $a_{min}$ values from 1.1 to 2.2 AU in the first column of the table

| $a_{min}$, AU | $e_0$ | $k_c$ | $N_2$ | $N_4$ | $N_6$ | $N_8$ | $N_{10}$ | $N_{15}$ | $N_{20}$ | $N_{>20}$ | $N_s$ | $i_{min}$ | $i_{max}$ |
|---|---|---|---|---|---|---|---|---|---|---|---|---|---|
| $i_n=$ | | | 0 | 2° | 4° | 6° | 8° | 10° | 15° | 20° | | | |
| $i_x=$ | | | 2° | 4° | 6° | 8° | 10° | 15° | 20° | — | | | |
| 1.1 | 0.02 | 1 | 2 | 2 | 2 | 2 | 0 | 0 | 0 | 0 | 8 | 1.86 | 7.76 |
| 1.2 | 0.02 | 1 | 14 | 20 | 24 | 13 | 6 | 6 | 2 | 0 | 85 | 0.22 | 19.6 |
| 1.3 | 0.02 | 1 | 19 | 24 | 24 | 24 | 12 | 11 | 0 | 0 | 114 | 0.34 | 14.1 |
| 1.4 | 0.02 | 1 | 16 | 33 | 26 | 28 | 6 | 9 | 1 | 1 | 120 | 0.51 | 22.1 |
| 1.4 | 0.02 | 1* | 16 | 20 | 29 | 12 | 7 | 8 | 0 | 0 | 92 | 0.50 | 14.3 |
| 1.4 | 0.02 | 1** | 10 | 31 | 18 | 17 | 6 | 10 | 1 | 1 | 94 | 0.05 | 21.8 |
| 1.4 | 0.02 | 1*** | 8 | 27 | 21 | 17 | 5 | 10 | 1 | 0 | 89 | 1.03 | 17.4 |
| 1.5 | 0.02 | 1 | 16 | 23 | 40 | 19 | 9 | 8 | 3 | 1 | 119 | 0.31 | 31.2 |
| 1.6 | 0.02 | 1 | 20 | 37 | 22 | 15 | 7 | 5 | 2 | 1 | 109 | 0.40 | 27.5 |
| 1.7 | 0.02 | 1 | 88 | 29 | 6 | 13 | 4 | 3 | 0 | 0 | 143 | 0.16 | 14.2 |
| 1.8 | 0.02 | 1 | 26 | 3 | 2 | 0 | 0 | 0 | 0 | 0 | 31 | 1.08 | 5.91 |
| $\Sigma_{ni}$ | 0.02 | 1 | 235 | 249 | 214 | 160 | 62 | 70 | 10 | 4 | 1004 | | |
| $\Sigma_{fi}$ | 0.02 | 1 | 0.234 | 0.248 | 0.213 | 0.159 | 0.062 | 0.07 | 0.010 | 0.004 | 1.0 | 0.05 | 31.2 |
| 1.0 | 0.15 | 1 | 4 | 14 | 11 | 5 | 6 | 1 | 1 | 0 | 42 | 1.17 | 18.3 |
| 1.1 | 0.15 | 1 | 10 | 15 | 22 | 25 | 9 | 11 | 4 | 2 | 98 | 0.78 | 22.2 |
| 1.2 | 0.15 | 1 | 21 | 36 | 38 | 29 | 19 | 14 | 3 | 0 | 160 | 0.23 | 17.2 |
| 1.3 | 0.15 | 1 | 12 | 40 | 37 | 27 | 19 | 15 | 8 | 0 | 158 | 0.99 | 19.6 |
| 1.3 | 0.15 | 1** | 25 | 34 | 33 | 27 | 16 | 21 | 5 | 1 | 162 | 0.29 | 23.6 |
| 1.4 | 0.15 | 1 | 19 | 33 | 35 | 30 | 27 | 34 | 4 | 1 | 183 | 0.22 | 20.4 |
| 1.4 | 0.15 | 1* | 18 | 38 | 35 | 34 | 21 | 27 | 1 | 2 | 176 | 0.06 | 21.3 |
| 1.4 | 0.15 | 1** | 20 | 41 | 34 | 28 | 27 | 24 | 8 | 0 | 182 | 0.34 | 18.6 |
| 1.4 | 0.15 | 1*** | 14 | 28 | 43 | 32 | 24 | 27 | 5 | 1 | 174 | 0.58 | 30.0 |
| 1.5 | 0.15 | 1 | 18 | 35 | 32 | 34 | 27 | 28 | 3 | 1 | 178 | 0.29 | 24.7 |
| 1.5 | 0.15 | 1** | 17 | 32 | 43 | 28 | 25 | 22 | 8 | 2 | 177 | 0.46 | 29.5 |
| 1.5 | 0.15 | 1*** | 16 | 43 | 33 | 32 | 17 | 22 | 6 | 3 | 172 | 0.51 | 26.1 |
| 1.6 | 0.15 | 1 | 20 | 39 | 43 | 37 | 14 | 21 | 6 | 0 | 180 | 0.69 | 19.6 |
| 1.7 | 0.15 | 1 | 17 | 30 | 37 | 36 | 13 | 18 | 4 | 1 | 156 | 0.20 | 23.5 |
| 1.8 | 0.15 | 1 | 11 | 31 | 27 | 25 | 17 | 11 | 2 | 1 | 125 | 0.68 | 20.8 |
| 1.9 | 0.15 | 1 | 12 | 29 | 27 | 23 | 7 | 6 | 0 | 1 | 105 | 0.35 | 21.8 |
| 2.0 | 0.15 | 1 | 8 | 109 | 87 | 22 | 9 | 3 | 0 | 0 | 238 | 1.03 | 11.4 |
| 2.1 | 0.15 | 1 | 23 | 95 | 55 | 22 | 11 | 8 | 2 | 1 | 217 | 0.61 | 21.9 |
| 2.2 | 0.15 | 1 | 10 | 31 | 25 | 9 | 2 | 0 | 1 | 0 | 78 | 0.22 | 19.9 |
| $\Sigma_{ni}$ | 0.15 | 1 | 291 | 739 | 686 | 499 | 304 | 315 | 70 | 17 | 2919 | | |
| $\Sigma_{fi}$ | 0.15 | 1 | 0.100 | 0.253 | 0.235 | 0.171 | 0.104 | 0.108 | 0.024 | 0.006 | 1.0 | 0.05 | 31.2 |
| 1.4 | 0.02 | 0.5 | 6 | 14 | 20 | 7 | 12 | 9 | 2 | 1 | 71 | 0.90 | 21.4 |
| 1.5 | 0.02 | 0.5 | 3 | 19 | 25 | 14 | 10 | 13 | 1 | 1 | 86 | 1.35 | 21.7 |

**Table 5.** (Contd.)

| $a_{min}$, AU | $e_0$ | $k_c$ | $N_2$ | $N_4$ | $N_6$ | $N_8$ | $N_{10}$ | $N_{15}$ | $N_{20}$ | $N_{>20}$ | $N_s$ | $i_{min}$ | $i_{max}$ |
|---|---|---|---|---|---|---|---|---|---|---|---|---|---|
| $\Sigma_{ni}$ | 0.02 | 0.5 | 9 | 33 | 45 | 21 | 22 | 22 | 3 | 2 | 157 | | |
| $\Sigma_{fi}$ | 0.02 | 0.5 | 0.057 | 0.210 | 0.287 | 0.134 | 0.140 | 0.140 | 0.019 | 0.013 | 1.0 | 0.90 | 21.7 |
| 1.4 | 0.15 | 0.5 | 15 | 28 | 27 | 24 | 13 | 12 | 3 | 5 | 127 | 0.14 | 30.5 |
| 1.5 | 0.15 | 0.5 | 13 | 30 | 48 | 28 | 23 | 20 | 3 | 1 | 166 | 0.42 | 20.7 |
| $\Sigma_{ni}$ | 0.15 | 0.5 | 26 | 58 | 75 | 52 | 36 | 32 | 6 | 6 | 293 | | |
| $\Sigma_{fi}$ | 0.15 | 0.5 | 0.089 | 0.198 | 0.226 | 0.177 | 0.123 | 0.109 | 0.020 | 0.020 | 1.0 | 0.14 | 30.5 |
| 1.4 | 0.02 | 0.1 | 1 | 4 | 2 | 1 | 0 | 1 | 0 | 0 | 9 | 0.69 | 10.3 |
| 1.5 | 0.02 | 0.1 | 1 | 8 | 3 | 2 | 1 | 0 | 0 | 2 | 17 | 1.38 | 11.7 |
| $\Sigma_{ni}$ | 0.02 | 0.1 | 2 | 12 | 5 | 3 | 1 | 1 | 0 | 2 | 26 | | |
| $\Sigma_{fi}$ | 0.02 | 0.1 | 0.077 | 0.462 | 0.192 | 0.115 | 0.038 | 0.038 | 0 | 0.077 | 1.0 | 0.69 | 11.7 |
| 1.4 | 0.15 | 0.1 | 10 | 8 | 2 | 3 | 5 | 3 | 0 | 4 | 35 | 0.06 | 25.3 |
| 1.5 | 0.15 | 0.1 | 7 | 21 | 8 | 7 | 5 | 4 | 1 | 2 | 55 | 0.05 | 39.5 |
| $\Sigma_{ni}$ | 0.15 | 0.1 | 17 | 29 | 10 | 10 | 9 | 8 | 1 | 6 | 90 | | |
| $\Sigma_{fi}$ | 0.15 | 0.1 | 0.189 | 0.322 | 0.111 | 0.111 | 0.1 | 0.089 | 0.011 | 0.067 | 1.0 | 0.05 | 39.5 |

$N_2$, $N_4$, $N_6$, $N_8$, $N_{10}$, $N_{15}$, $N_{20}$, $N_{>20}$ represent the number of planetesimals whose orbital inclinations fall within the intervals ($i_n$, $i_x$) from $0°$ to $2°$, from $2°$ to $4°$, from $4°$ to $6°$, from $6°$ to $8°$, from $8°$ to $10°$, from $10°$ to $15°$, from $15°$ to $20°$, and greater than $20°$, respectively. $N_s$ is the total number of planetesimals that reached 500 AU from the star. $i_{min}$ and $i_{max}$ are the minimum and maximum orbital inclinations in degrees. In each calculation variant, 250 planetesimals with orbital semimajor axes from $a_{min}$ to $a_{min}$ + 0.1 AU, eccentricities $e_0$ and inclinations $e_0/2$ rad, respectively, were considered. The mass of the planet in the orbit of planet $c$ was $7k_c m_E$. $\Sigma_{ni}$ is the sum of the numbers of planetesimals in the column in the variants presented above. $S_{fi} = \Sigma_{ni}/N_s$.

from 500 to 1200 AU equal to 1 million years, the approximate total mass of planetesimals at distances of 500 to 1200 AU was $0.5m_E$. The time for the motion of such planetesimals at distances less than 500 AU from the star was, on average, less than the time of their motion at distances from 500 to 1200 AU. Accounting for this motion slightly reduces the estimate of the mass of matter in the zone from 500 to 1200 AU. After the main cleaning of the planet $c$'s feeding zone from planetesimals, the number of planetesimals at distances of 500–1200 AU was much less than during the accumulation of planet $c$. For some time, the ejection of planetesimals could occur from the outer part of the planet $c$'s feeding zone. Individual planetesimals could be ejected from this part of the zone after hundreds of millions of years. Within the framework of the considered model (which only took into account the migration of planetesimals from the feeding zone of planet $c$), the zone from 500 to 1200 AU from the star became almost empty after tens of millions of years.

Due to mutual gravitational interactions and collisions, some of the planetesimals could increase the perihelion distances of their orbits and subsequently maintained their orbits nearly unchanged due to interactions with planet $c$. These planetesimals could form a low-mass analog of the Hills cloud. It is possible that beyond the orbit of planet $c$ at the stage of planet formation, there were many planetesimals formed from a part of the gas and dust disk. Interactions between planetesimals from the feeding zone of planet $c$, moving in highly eccentric orbits, and planetesimals formed outside the feeding zone of planet $c$, could increase the influx of bodies into the analog of the Hills cloud.

Currently, the region between between 500 and 1200 AU is likely to be nearly empty if we consider its replenishment only by planetesimals from the feeding zone of planet $c$. Planetesimals ejected from the feeding zone of planet $c$ had relatively small orbital inclinations. Even colliding and gravitationally interacting with each other, they could not significantly increase the average inclination of their orbits. Large inclinations of the orbits of bodies are likely to be primarily associated with the bodies that came from outside into the Hill sphere of the star Proxima Centauri. The radius of the Hill sphere of Proxima Centauri is an order of magnitude smaller than the outer boundary radius of the Hills cloud in the Solar System and two orders of magnitude smaller than the radius of the Hill sphere of the Sun. Therefore, it is difficult to expect the existence of a similarly massive cloud around this star as the Oort cloud around the Sun. Some authors (Levison et al., 2010; Siraj and Loeb, 2021) suggest that most of the bodies in the Oort cloud of the Solar System have an extrasolar origin. It seems unlikely that

**Table 6.** Distribution of planetesimals that first reached 1200 AU from the star by the orbital inclination intervals $(i_n, i_x)$ for $e_0$ equal to 0.02 or 0.15 and several $a_{min}$ values from 1.1 to 2.2 AU given in the first column of the table

| $a_{min}$, AU | $e_0$ | $k_c$ | $N_2$ | $N_4$ | $N_6$ | $N_8$ | $N_{10}$ | $N_{15}$ | $N_{20}$ | $N_{>20}$ | $N_s$ | $i_{min}$ | $i_{max}$ |
|---|---|---|---|---|---|---|---|---|---|---|---|---|---|
| $i_n=$ | | | 0 | 2° | 4° | 6° | 8° | 10° | 15° | 20° | | | |
| $i_x=$ | | | 2° | 4° | 6° | 8° | 10° | 15° | 20° | — | | | |
| 1.2 | 0.02 | 1 | 14 | 19 | 23 | 16 | 4 | 5 | 3 | 1 | 85 | 0.19 | 21.2 |
| 1.3 | 0.02 | 1 | 21 | 25 | 25 | 17 | 15 | 10 | 1 | 0 | 114 | 0.30 | 16.5 |
| 1.4 | 0.02 | 1 | 16 | 32 | 26 | 22 | 7 | 12 | 0 | 2 | 117 | 0.47 | 24.9 |
| 1.5 | 0.02 | 1 | 17 | 23 | 37 | 22 | 11 | 4 | 4 | 1 | 119 | 0.31 | 30.6 |
| 1.6 | 0.02 | 1 | 20 | 36 | 21 | 13 | 8 | 7 | 1 | 3 | 109 | 0.45 | 27.1 |
| 1.7 | 0.02 | 1 | 85 | 30 | 7 | 14 | 4 | 2 | 1 | 0 | 143 | 0.16 | 17.1 |
| $\Sigma_{ni}$ | 0.02 | 1 | 173 | 165 | 139 | 104 | 49 | 40 | 10 | 7 | 687 | 0.16 | |
| $\Sigma_{fi}$ | 0.02 | 1 | 0.252 | 0.240 | 0.202 | 0.151 | 0.071 | 0.058 | 0.015 | 0.010 | 1.0 | 0.16 | 30.6 |
| 1.1 | 0.15 | 1 | 8 | 18 | 22 | 19 | 11 | 12 | 4 | 2 | 96 | 0.85 | 21.0 |
| 1.2 | 0.15 | 1 | 20 | 36 | 39 | 30 | 17 | 15 | 3 | 0 | 160 | 0.24 | 17.5 |
| 1.3 | 0.15 | 1 | 14 | 34 | 38 | 32 | 16 | 17 | 6 | 1 | 158 | 0.98 | 20.8 |
| 1.4 | 0.15 | 1 | 19 | 31 | 32 | 30 | 37 | 25 | 9 | 0 | 183 | 0.22 | 18.8 |
| 1.5 | 0.15 | 1 | 19 | 32 | 35 | 37 | 22 | 26 | 4 | 3 | 178 | 0.26 | 27.6 |
| 1.5 | 0.15 | 1** | 15 | 37 | 36 | 31 | 29 | 19 | 9 | 1 | 177 | 0.46 | 34.4 |
| 1.5 | 0.15 | 1*** | 16 | 42 | 35 | 29 | 22 | 20 | 5 | 3 | 172 | 0.56 | 28.5 |
| 1.6 | 0.15 | 1 | 18 | 38 | 52 | 27 | 21 | 17 | 5 | 2 | 180 | 0.66 | 22.4 |
| 1.7 | 0.15 | 1 | 18 | 32 | 37 | 31 | 15 | 19 | 3 | 1 | 156 | 0.21 | 21.0 |
| 1.8 | 0.15 | 1 | 11 | 30 | 20 | 28 | 15 | 15 | 2 | 3 | 124 | 0.66 | **179** |
| 1.9 | 0.15 | 1 | 12 | 27 | 29 | 21 | 10 | 5 | 0 | 1 | 105 | 0.337 | 22.0 |
| 2.0 | 0.15 | 1 | 7 | 109 | 84 | 24 | 10 | 3 | 0 | 0 | 237 | 1.101 | 11.4 |
| 2.1 | 0.15 | 1 | 22 | 86 | 64 | 24 | 12 | 7 | 3 | 1 | 219 | 0.535 | **48.8** |
| 2.2 | 0.15 | 1 | 13 | 33 | 22 | 11 | 3 | 1 | 1 | 0 | 84 | 0.23 | 17.4 |
| $\Sigma_{ni}$ | 0.15 | 1 | 212 | 615 | 545 | 374 | 240 | 201 | 54 | 18 | 2229 | | |
| $\Sigma_{fi}$ | 0.15 | 1 | 0.095 | 0.276 | 0.245 | 0.168 | 0.108 | 0.090 | 0.024 | 0.008 | 1.0 | 0.21 | 179 |
| 1.4 | 0.02 | 0.5 | 7 | 11 | 19 | 10 | 9 | 11 | 3 | 1 | 71 | 0.87 | 23.2 |
| 1.5 | 0.02 | 0.5 | 4 | 19 | 25 | 14 | 8 | 13 | 2 | 1 | 86 | 1.28 | 24.1 |
| $\Sigma_{ni}$ | 0.02 | 0.5 | 11 | 30 | 44 | 24 | 17 | 24 | 5 | 2 | 157 | | |
| $\Sigma_{fi}$ | 0.02 | 0.5 | 0.070 | 0.191 | 0.280 | 0.153 | 0.108 | 0.153 | 0.032 | 0.013 | 1.0 | 0.87 | 24.1 |
| 1.4 | 0.15 | 0.5 | 15 | 26 | 27 | 28 | 9 | 15 | 4 | 4 | 128 | 0.13 | 26.5 |
| 1.5 | 0.15 | 0.5 | 12 | 32 | 47 | 27 | 23 | 17 | 5 | 2 | 165 | 0.41 | 20.9 |
| $\Sigma_{ni}$ | 0.15 | 0.5 | 27 | 58 | 74 | 55 | 32 | 32 | 9 | 6 | 293 | 0.13 | 26.5 |
| $\Sigma_{fi}$ | 0.15 | 0.5 | 0.092 | 0.198 | 0.253 | 0.188 | 0.109 | 0.109 | 0.031 | 0.020 | 1.0 | 0.13 | 26.5 |
| 1.4 | 0.02 | 0.1 | 3 | 2 | 2 | 1 | 1 | 0 | 0 | 0 | 9 | 0.29 | 9.51 |
| $\Sigma_{fi}$ | 0.02 | 0.1 | 0.33 | 0.22 | 0.22 | 0.11 | 0.11 | 0 | 0 | 0 | 1.0 | | |
| 1.4 | 0.15 | 0.1 | 9 | 7 | 3 | 5 | 6 | 1 | 0 | 4 | 35 | 0.07 | **175** |
| $\Sigma_{fi}$ | 0.15 | 0.1 | 0.26 | 0.2 | 0.086 | 0.143 | 0.171 | 0.029 | 0 | 0.114 | 1.0 | 0.07 | 175 |

$N_2$, $N_4$, $N_6$, $N_8$, $N_{10}$, $N_{15}$, $N_{20}$, $N_{>20}$ represent the number of planetesimals whose orbital inclinations fall within the intervals $(i_n, i_x)$ from 0° to 2°, from 2° to 4°, from 4° to 6°, from 6° to 8°, from 8° to 10°, from 10° to 15°, from 15° to 20°, and greater than 20°, respectively. $N_s$ is the total number of planetesimals that reached 1200 AU from the star. $i_{min}$ and $i_{max}$ are the minimum and maximum orbital inclinations in degrees. In each calculation variant, 250 planetesimals with orbital semimajor axes from $a_{min}$ to $a_{min}$ + 0.1 AU, eccentricities $e_0$ and inclinations $e_0/2$ rad, respectively, were considered at the initial time. The mass of the planet in the orbit of planet $c$ was $7k_c m_E$. $\Sigma_{ni}$ is the sum of the numbers of planetesimals in the column for the variants presented above. $S_{fi} = \Sigma_{ni}/N_s$.

**Table 7.** Fraction of planetesimals with orbital inclinations greater than 10° among all planetesimals that first reached a distance from the star equal to $a_{lim}$

| $a_{lim}$, AU | 500 | 500 | 500 | 1200 | 1200 | 1200 |
|---|---|---|---|---|---|---|
| $k_c$ | 1 | 0.5 | 0.1 | 1 | 0.5 | 0.1 |
| $e_0 = 0.02$ | 0.08 | 0.17 | 0.11 | 0.08 | 0.20 | 0 |
| $e_0 = 0.15$ | 0.14 | 0.15 | 0.17 | 0.12 | 0.16 | 0.14 |

**Table 8.** Fraction of planetesimals that increased their maximum distance from the star from 500 to 1200 AU over the time interval d$t$ for calculation variants with different $a_{min}$ value at $e_0 = 0.02$ and $t_s = 1$ day

| $a_{min}$, AU | 1.2 | 1.3 | 1.4 | 1.5 | 1.6 | 1.7 |
|---|---|---|---|---|---|---|
| d$t$ < 0.01 Myr | 0.188 | 0.096 | 0.143 | 0.109 | 0.046 | 0.098 |
| d$t$ < 0.1 Myr | 0.529 | 0.333 | 0.487 | 0.487 | 0.321 | 0.371 |
| $0.1 \le$ d$t$ < 1 Myr | 0.307 | 0.492 | 0.429 | 0.353 | 0.578 | 0.503 |
| $1 \le$ d$t$ < 2 Myr | 0.082 | 0.114 | 0.0 | 0.084 | 0.046 | 0.077 |
| d$t$ > 2 Myr | 0.082 | 0.061 | 0.084 | 0.076 | 0.055 | 0.049 |
| d$t$ > 5 Myr | 0.035 | 0.026 | 0.017 | 0.034 | 0.018 | 0.028 |
| d$t$ > 10 Myr | 0.012 | 0.017 | 0 | 0.025 | 0.018 | 0.007 |
| d$t_{max}$, Myr | 59.76 | 23.48 | 5.12 | 18.80 | 24.56 | 11.52 |

d$t_{max}$ is the maximum value of d$t$.

Proxima Centauri, with its relatively small Hill sphere, could capture bodies coming from the vicinity of other stars.

The results of the calculations (in particular, the distribution of planetesimal orbits in the outer part of the Hill sphere by their inclinations and eccentricities) can be used as initial data for models that take into account gravitational interactions and collisions of planetesimals, leading to the formation of an analog of the Hills cloud for Proxima Centauri. These results can also serve as initial data in models that consider the gravitational influence of a neighboring pair of stars to estimate its impact on the evolution of plane-tesimal orbits in the outer part of the Hill sphere and the possibility of returning individual ejected planetesimals back to the Hill sphere, but with more inclined orbits around the star Proxima Centauri.

There may be analogs of asteroid and trans-Neptunian belts around Proxima Centauri, and they could be more numerous than in the Solar System. The lower mass ratio of planet $c$ to the star, compared to Jupiter, larger semimajor axis ratios of the orbits of planets $c$ and $b$, compared to Jupiter and Mars, and the presence of only one major planet in the Proxima Centauri system could be reasons for such differences in the belts.

**Table 9.** Fraction of planetesimals that increased their maximum distance from the star from 500 to 1200 AU over the time interval d$t$ for disks with different $a_{min}$ and $t_s$ values at $e_0 = 0.15$

| $a_{min}$, AU | 1.2 | 1.3 | 1.4 | 1.5 | 1.5 | 1.5 | 1.6 | 1.7 |
|---|---|---|---|---|---|---|---|---|
| $t_s$, days | 1 | 1 | 1 | 0.5 | 1 | 2 | 1 | 1 |
| d$t$ < 0.01 Myr | 0.181 | 0.234 | 0.213 | 0.237 | 0.219 | 0.273 | 0.233 | 0.212 |
| d$t$ < 0.1 Myr | 0.456 | 0.538 | 0.519 | 0.458 | 0.545 | 0.430 | 0.578 | 0.436 |
| $0.1 \le$ d$t$ < 1 Myr | 0.45 | 0.392 | 0.443 | 0.497 | 0.343 | 0.518 | 0.394 | 0.481 |
| $1 \le$ d$t$ < 2 Myr | 0.063 | 0.051 | 0.027 | 0.028 | 0.056 | 0.017 | 0.011 | 0.045 |
| d$t$ > 2 Myr | 0.031 | 0.019 | 0.011 | 0.011 | 0.056 | 0.035 | 0.017 | 0.038 |
| d$t$ > 5 Myr | 0.006 | 0.006 | 0 | 0 | 0.0225 | 0.017 | 0.006 | 0.013 |
| d$t$ > 10 Myr | 0 | 0 | 0 | 0 | 0.0112 | 0.0116 | 0.006 | 0.006 |
| d$t_{max}$, Myr | 5.66 | 6.71 | 3.639 | 3.15 | 29.36 | 11.47 | 10.78 | 17.5 |

d$t_{max}$ is the maximum value of d$t$.

**Table 10.** Fraction of planetesimals that increased their maximum distance from the star from 500 to 1200 AU over the time interval d$t$ for disks with different $a_{min}$ and $e_0$ values and ratio $k_c$ of the planet's mass to the present mass of planet $c$

| $a_{min}$, AU | 1.2−1.7 | 1.2−1.7 | 1.2−1.7 | 1.4 | 1.5 | 1.4 | 1.5 |
|---|---|---|---|---|---|---|---|
| $e_0$ | 0.02 | 0.15 | 0.15 | 0.02 | 0.02 | 0.15 | 0.15 |
| $k_c$ | 1 | 1 | 1 | 0.5 | 0.5 | 0.5 | 0.5 |
| $t_s$, days | 1 | 1 | 0.5−2 | 1 | 1 | 1 | 1 |
| d$t$ < 0.01 Myr | 0.113 | 0.215 | 0.225 | 0.114 | 0.072 | 0.134 | 0.108 |
| d$t$ < 0.1 Myr | 0.421 | 0.520 | 0.501 | 0.229 | 0.229 | 0.276 | 0.265 |
| 0.1 ≤ d$t$ < 1 Myr | 0.444 | 0.417 | 0.440 | 0.514 | 0.494 | 0.519 | 0.476 |
| 1 ≤ d$t$ < 2 Myr | 0.067 | 0.042 | 0.037 | 0.086 | 0.120 | 0.079 | 0.132 |
| d$t$ > 2 Myr | 0.068 | 0.029 | 0.027 | 0.171 | 0.157 | 0.126 | 0.127 |
| d$t$ > 5 Myr | 0.026 | 0.009 | 0.009 | 0.057 | 0.060 | 0.063 | 0.054 |
| d$t$ > 10 Myr | 0.013 | 0.004 | 0.004 | 0.014 | 0.012 | 0.024 | 0.024 |
| d$t_{max}$, Myr | 59.76 | 29.36 | 29.36 | 21.54 | 31.41 | 18.36 | 51.20 |

d$t_{max}$ is the maximum value of d$t$. For columns 1.2−1.7, the fractions are the averages for $a_{min}$ from 1.2 to 1.7 AU.

## CONCLUSIONS

This study examined the motion of planetesimals initially located in the feeding zone of planet Proxima Centauri $c$ at distances from the star from 500 to 1200 AU. The latter distance corresponds to the radius of the star's Hill sphere. The integration of the planetesimals' motion took into account the gravitational influence of the star and planets $b$ and $c$, as well as collisions between planetesimals and planets or the star. Most planetesimals from the larger part of the feeding zone of the nearly formed planet $c$ first reached 500 AU from the star within the first 10 million years. Only for planetesimals initially located at the edges of the feeding zone, more than half of them take longer than 10 million years to reach 500 AU. Individual planetesimals could reach the outer part of Proxima Centauri's Hill sphere even after hundreds of millions of years.

Approximately 90% of the planetesimals that first reached 500 AU from Proxima Centauri first reached 1200 AU from the star in less than 1 million years at the current mass of planet $c$. At the same time, no more than 2% of the planetesimals that reached 500 AU had aphelion distances between 500 and 1200 AU for more than 10 million years (but less than several tens of millions of years). When the mass of the planet was half the mass of planet $c$, the fraction of planetesimals that increased their maximum distance from the star from 500 to 1200 AU in less than 1 million years was about 70−80%.

For planetesimals that first reached 500 AU from the star at the current mass of planet $c$, the fraction of planetesimals with orbital eccentricities greater than 1 was 0.05 and 0.1 at the initial eccentricities $e_0 = 0.02$ and $e_0 = 0.15$, respectively. Among the planetesimals that first reached 1200 AU from the star, this fraction was about 0.3 for both $e_0$ values. The minimum values of the orbital eccentricity for planetesimals that reached 500 and 1200 AU were 0.992 and 0.995, respectively. The distribution of planetesimals that reached 500 AU from the star in terms of orbital eccentricities did not differ significantly at different times during planet $c$'s accretion (for different values of the growing planet's mass).

In the considered model, the disk of planetesimals in the outer part of the Hill sphere was rather flat. Inclinations $i$ of the orbits for over 80% of the planetesimals that first reached 500 or 1200 AU from the star did not exceed 10°. The fraction of such planetesimals with $i > 20°$ in most series of calculations did not exceed 2%. On average across all calculation variants, given the current mass of planet $c$, this fraction did not exceed 1%.

The motion of planetesimals was studied within (mostly deep inside) the Hill sphere of Proxima Centauri, and the ejected planetesimals had very little chance of returning to the Hill sphere of the star. Therefore, taking into account the gravitational influence of the binary star system (Alpha Centauri AB) is unlikely to change the conclusions drawn in this study. Strongly inclined orbits of bodies in the outer part of the Hill sphere of Proxima Centauri are likely due to objects coming from outside.

The results may be of interest for understanding the motion of bodies in some other exoplanetary systems, especially those with a dominant planet. They can be used to provide the initial data for models of the evolution of the disk of bodies in the outer part of Proxima Centauri's Hill sphere, which take into account gravitational interactions and collisions between bodies, as well as the influence of other stars.

The radius of the Hill sphere of Proxima Centauri is an order of magnitude smaller than the outer boundary radius of the Hills cloud in the Solar System and two orders of magnitude smaller than the Hill sphere radius of the Sun. Therefore, it is difficult to expect the existence of a similarly massive cloud around this star as the Oort cloud around the Sun.

## ACKNOWLEDGMENTS

The author expresses his deep gratitude to the referee for useful remarks that contributed to the improvement of the paper.

## FUNDING

## CONFLICT OF INTEREST

The authors of this work declare that they have no conflicts of interest.